\documentclass[prl,twocolumn,showpacs,superscriptaddress,amssymb]{revtex4}
\usepackage{graphicx}
\usepackage{bm}
\usepackage{dcolumn}
\newcommand{\barra}[1]{\,\overline{{\mathrm{#1}}}\,}
\begin{document}
\title{Surface Reactivity and Quantum-Size effects on the Electronic Density 
 Decay Length of ultrathin Metal Films
}
\author{N. Binggeli}
\affiliation{
The Abdus Salam International Center for Theoretical 
Physics, Trieste 34014 , Italy} 
\affiliation{INFM-CNR DEMOCRITOS National Simulation Center, 
Trieste 34014 , Italy}
\author{M. Altarelli} 
\affiliation{
The Abdus Salam International Center for Theoretical 
Physics, Trieste 34014 , Italy} 
\affiliation{
European XFEL Project Team, DESY, Notkestra\ss e 85, 22607, Hamburg, Germany}
%
%
\begin{abstract}
The origin of the correlation between surface reactivity and 
quantum-size effects, observed in recent experiments on the oxidation 
of ultrathin magnesium films, is addressed by means of 
{\it ab initio} calculations and model predictions. We show 
that the decay length in vacuum of the electronic local density of states 
at the Fermi energy exhibits systematic oscillations with film 
thickness, with  local maxima induced when a quantum well state at 
$k_{\parallel}=0$ crosses the Fermi energy. The predicted changes in 
the decay length are expected to have a major impact on the 
electron transfer rate by tunneling, which has been proposed to control the 
initial sticking of O$_2$ in the oxidation process. 

\end{abstract}
\pacs{73.20.At, 73.21.Fg, 82.65.+r, 68.43.-h}

\maketitle
There is considerable interest in identifying methods to tailor the chemical 
reactivity of surfaces, a crucial factor in many technologically relevant 
surface phenomena, including  oxidation and catalysis. The recent 
observation of a correlation between quantum-size effects and the surface 
reactivity of ultrathin metal films \cite{Aballe04} is an exciting development 
in this area, both because of its fundamental interest for a quantitative 
understanding of the structure-size dependence of the reactivity and in view 
of the importance of nano-scale and low-dimensional structures in modern 
technology. 
Experimentally, oscillations were observed in the oxidation rate 
of ultrathin Mg(0001) films on W(110), as a function of film 
thickness, with the largest oxidation rate occurring when a quantum-well 
state was found to cross the Fermi energy, in photoemission (PE) 
spectra \cite{Aballe04} taken near normal incidence. 
In particular, the changes observed in the initial oxidation rate---when 
most of the film was still metallic, were found to be dramatic. 
The origin of these changes has not been established yet.  
The changes in reactivity were suggested to be due to a change 
in the density of states (DOS) at, or near the Fermi energy $E_F$ \cite{Aballe04}, 
as the oscillations in the reactivity were found to correlate with the 
oscillations in the PE intensity at $E_F$. 
However, unlike the normal-incidence PE intensity, which measures 
the partial DOS at $k_{\parallel} = 0$, the total DOS of the films 
at $E_F$ is bound to exhibit a monotonic, staircase-like increase 
with film thickness \cite{Boettger}, and cannot thus 
simply account for the oscillations in the reactivity. 

Here we propose a theoretical interpretation of these observations, 
in which the decay length in the vacuum of the electronic local density of 
states around the Fermi energy is identified as the key parameter 
responsible for the changes in the reactivity. On the basis of 
{\it ab initio} calculations, 
inspired by model predictions for quantum well states, 
we show that the decay length, $\lambda$, of the local density of states 
at the Fermi energy exhibits an oscillatory behavior with film 
thickness, with local maxima present when the highest occupied 
quantum-well state at  $k_{\parallel} = 0$  is closest to the Fermi 
energy. The predicted changes in the decay length  are significant. 
They  are expected to have a direct, exponential impact on the electron 
transfer rate by tunneling, which has been proposed to control the initial 
sticking of O$_2$  on the metal surface \cite{Hellman03}. 
The changes in $\lambda$ should be  observable 
by scanning tunneling microscopy (STM), near terrace steps, on surfaces 
characterized by regions of different film thicknesses \cite{Aballe04}. 

{\em Model predictions for $\lambda$.}---Within an independent electron 
description and considering, for simplicity, a square-well potential along 
the  $z$ direction, normal to the thin film, with a constant depth $V$ and 
a variable width $L$ (film thickness), the electronic states are  
solutions of a separable problem in the $z$ and $(x,y)$ 
variables,  and read: 
\begin{equation}
\psi^{E}_{n,k_x,k_y} (x,y,z) \sim 
\chi_n (z) \cdot e^{i(k_x \cdot x + k_y \cdot y)},  
\end{equation}
with energy $E = E_n + \hbar^2 (k_x^2 + k_y^2) / 2 m^{*}$, where  $m^{*}$ 
stands for the electron effective mass. 
$E_n$ and $\chi_n (z)$ are the 
eigenvalues and eigenstates of the one-dimensional square-well potential 
problem, with the following properties: 
\begin{equation}
\chi_n(z) \sim e^{-\alpha_n \cdot z} 
\end{equation}
for $z \geq L/2$, assuming the $z$ origin at the center of the film, and:  
\begin{equation}
\alpha_n = (\sqrt{2m^{*}}/\hbar) \cdot \sqrt{-E_n},  
\end{equation} 
where the zero of energy is taken at the vacuum level.  In the three-dimensional 
problem, $E_n$ coincides with the energy $E$ of the subband state $n$ at 
$k_{\parallel} = 0$, measured relative to the vacuum level. 

All states $\psi^{E}_{n,k_x,k_y}$ belonging to subband $n$, with energy 
$E \geq E_{n}$, are thus characterized by the same decay length 
$1/\alpha_{n} \sim 1/\sqrt{-E_n}$, for a given film thickness $L$. 
If $E_F$ is located between the levels $E_n$ and $E_{n+1}$,   
the dominant decay length of the electronic 
states at $E_F$ is $1/\alpha_{n}$, namely the decay length 
of the highest occupied band state at $k_{\parallel} = 0$. With increasing 
width $L$ of the well, the energies of the quantum-well states 
decrease with respect to $E_F$; hence, the decay length $\lambda$ 
of the electronic density at $E_F$ first decreases as 
$1/2\alpha_{n} \sim 1/\sqrt{-E_n}$, until the next quantum-well state 
at  $\barra{\Gamma}$ crosses  $E_F$, at which point 
$\lambda$  increases to the new value $1/2\alpha_{n+1} \sim 1/\sqrt{-E_{n+1}}$. 
The decay length then decreases again with increasing $L$, displaying 
systematic oscillations with $L$. 
Considering next a discrete number of atomic layers, the periodicity of the 
crossing of the Fermi energy may be derived from the Bohr-Sommerfeld rule, 
which for Mg(0001) yields a periodicity of 7.7 monolayers (ML) \cite{Schiller04}. 

From the model description, we expect thus oscillations 
in $\lambda$, with local maxima occurring when the highest occupied quantum-well 
states at $\barra{\Gamma}$ is closest to $E_F$. 
A more realistic description of the system, however, is clearly needed to 
confirm the robustness of this behavior, and to quantify the effect.  
In particular, electron interaction, atomic orbitals and bonds, surface 
states, and the presence of the substrate, are all factors which are expected 
to influence the decay length $\lambda$. We have therefore examined the effect 
of quantum confinement on $\lambda$ by means of {\it ab initio} calculations 
performed for epitaxial Mg(0001) films on W(110) and also, for comparison, for 
the corresponding unsupported Mg(0001) films in vacuum. 

{\em First-principles results.}---The calculations were performed within 
density functional theory, using the Perdew-Burke-Ernzerhof exchange-correlation 
functional \cite{pbe},  Troullier-Martins pseudopotentials \cite{TM},  
and a plane-wave basis set.  
The Mg pseudopotential was generated in the atomic   
configuration: $3s^{1.8} 3p^{0} 3d^{0.2}$, using as core-radii 
cutoff:  $r_{\rm s,p,d} = 2.8$ a.u.. For W we 
treated the $6s$, $6p$, and $5d$ orbitals as valence states, using the 
same parameters as in Ref.~\cite{Kwak96}. 
The overlap between valence and core electrons was accounted for    
using the non-linear-core correction to the exchange-correlation 
potential \cite{nlcc}. A kinetic-energy cutoff of $49$ Ry was used for 
the plane-wave expansion of the electronic orbitals of the Mg/W systems. 
For the isolated Mg(0001) slabs we used a kinetic-energy cutoff of 14 Ry. 
The Brillouin-zone sampling was done with a Monkhorst-Pack (MP) 
grid \cite{MP}, and we used a Gaussian smearing of the electronic levels 
of $0.02$ Ry to determine the Fermi energy. 
The films were modeled using slab geometries in supercells. 
For the Mg films on W, we considered slabs containing 7 W(110) 
ML, terminated by 5 to 12 Mg(0001) ML on one side, and by 2 
Mg (0001) ML on the other side.  The Mg bilayer was introduced in order 
to avoid the presence of an electric field---due to the different work 
functions of W and Mg---in the vacuum regions separating the periodic 
images of the slab. We used vacuum regions with a minimal thickness of 
20.1 \AA, in order to obtain well-converged values of $\lambda$. 

Experimentally, above 3 ML, the Mg films are known to grow epitaxially 
on W(110) \cite{Schiller04}, with lattice parameters corresponding to their 
bulk values \cite{Aballe}.  
The mismatches between the experimental in-plane lattice parameters of Mg(0001) 
($a_{\text{Mg}}=3.21$~\AA, $b_{\text{Mg}}=\sqrt{3} a_{\text{Mg}}$) and 
W(110) ($a_{\text{W}}=3.16$~\AA, $b_{\text{W}}=\sqrt{2} a_{\text{W}}$) 
are 1.6\,\% and $\sim20$\,\% \cite{Schiller04}. 
To model such epitaxial systems, a commensurate interface atomic structure 
is needed in the calculations. The inclusion of dislocations would require 
prohibitively large lateral dimensions of the supercell, and the details of 
the atomic structure at the interface are unknown in any case. In order to  
simulate an unstrained Mg film on W(110), we thus elected to laterally 
strain the W(110) slab to the in-plane lattice parameters of Mg(0001). 
The epitaxial alignment was made by positioning the atoms of the first W(110) 
layer, adjacent to the Mg, in the continuation of the Mg (0001) 
hcp lattice. 
The three outermost layers of the Mg film were relaxed, while the other 
interlayer spacings were kept frozen at their 
bulk position (using the theoretical values of the bulk lattice 
constants: $a_{\text{Mg}}=3.19$~\AA, $c_{\text{Mg}} = 5.18$~\AA, and 
$a_{\text{W}} = 3.21$~\AA). It should be noted, however, that the  Mg(0001) 
surface relaxations are small \cite{Staikov99,nous}, and were found to have 
a negligible impact on $\lambda$. 
The self-consistent calculations were carried out using a (20,20,1) MP grid. 
For the local density of states, we used  a (48,48,1) grid centered at 
$\Gamma$, with a Gaussian level smearing of $0.005$ Ry. The decay length 
was derived from a fit, assuming an exponential decay of the local density of 
states at distances beyond $\sim 2.15$~\AA\ from the outermost atomic plane.

\begin{figure}
  \includegraphics[width=6.4cm]{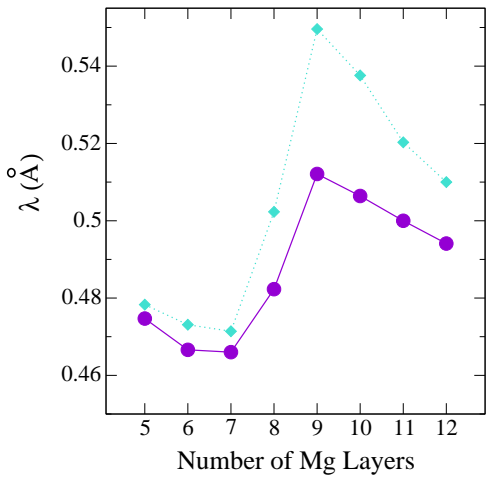}
    \caption[lambda]{\label{fig:lambda}
      (Color online) Calculated decay length in vacuum of the electronic local 
      density of 
      states at the Fermi energy of Mg(0001) films of various thicknesses 
      on W(110) (disks). The decay length of the corresponding Mg(0001) films 
      isolated in vacuum is also shown for comparison (diamonds).}
\end{figure}

\begin{figure}
\includegraphics[width=6.8cm]{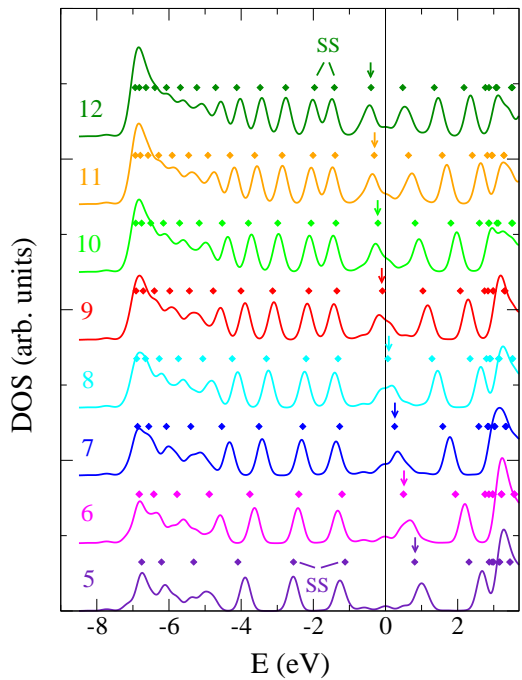}
    \caption[ldos]{\label{fig:ldos}
      (Color online) Calculated density of states at $\barra{\Gamma}$ of  
      the Mg (0001) films on W(110) (solid lines). The film thickness  
      increases from 5 to 12 atomic layers (bottom to top curve). The densities of 
      states have been convoluted with a Gaussian of width 0.2 eV. The symbols  
      (diamonds) show the energies of the electronic levels at $\barra{\Gamma}$ 
      of the isolated Mg(0001) films in vacuum. The states indicated by ``SS'' 
      originate from the Shockley surface state of Mg(0001) (see text). The zero 
      of energy corresponds to the Fermi level.}
\end{figure}

In Fig.~\ref{fig:lambda}, we report the calculated decay length $\lambda$ 
of the Mg films, as a function of film thickness. The results are shown 
for both the Mg films on W and the isolated Mg films in vacuum. The 
behaviors are very similar in the two cases. The decay length exhibits a 
pronounced oscillation, with a maximum at 9 layers and a minimum at 6-7 
layers. The presence of the tungsten substrate tends to reduce the amplitude 
of the variation  of $\lambda$, from 17\% to 10\%, but has no significant impact  
on the position of the extrema. 
The densities of states at $\barra{\Gamma}$ of the   
Mg (0001) films on W(110) are displayed in Fig.~\ref{fig:ldos}.  With increasing  
film thickness, an unoccupied  quantum-well state crosses the 
Fermi energy  at $\sim 9$ layers. This coincides with the maximal decay 
length $\lambda$ found in Fig.~\ref{fig:lambda}, consistent with the model 
prediction. 
In Fig.~\ref{fig:ldos}, we also reported  the energy positions of the 
electronic states at $\barra{\Gamma}$ obtained from the calculations 
for the isolated Mg slabs in vacuum. 
Also in this case, a quantum-well state is found to pass through the Fermi 
energy at  $\sim 9$ layers, which corresponds to the largest  $\lambda$  
obtained in Fig.~\ref{fig:lambda}. 
We note that we have also examined the variation of the workfunction of the 
Mg films on W with the number of Mg layers \cite{nousw}, as quantum 
size effects can be expected on this value \cite{Paggel02}. 
For the Mg films on W, however, the calculated changes in the work function 
are found to be very small, namely a variation of 0.05 eV in the range 5-12 
layers. With increasing thickness, the workfunction first 
decreases, in the range 5-8 layers, from 3.74 to 3.69 eV, and then  
increases, in the range 8-10 layers, from 3.69 to 3.72 eV,  and then saturates  
at 3.72 eV for higher coverage (in good agreement with the 
experimental value of 3.66 eV for Mg \cite{Michaelson77}). 

Inspection of Fig.~\ref{fig:ldos}  reveals a striking correspondence 
between the positions of the peaks in the DOS of 
the Mg films on W and the positions of the levels of the 
isolated Mg films in vacuum.  The largest difference between 
the two sets of energies  does not exceed 0.2 eV in the range 
[$E_F -8$ eV, $E_F +2$ eV]. 
The states indicated by ``SS'' in Fig.~\ref{fig:ldos} originate from the 
Shockley surface state of Mg(0001).  In the isolated Mg(0001) slab  
in vacuum, the Shockley states of the two surfaces  strongly 
interact, giving rise to a pair of split even and odd states---with respect to 
the center-of-slab reflection plane. The resulting splitting is 
as large as 1.4 eV for 5 Mg layers and 0.6 eV for 12 Mg layers.   
In the presence of the W substrate, these states persist as strong 
surface/interface resonances, with maxima in the probability density on 
both the outermost (surface) Mg layer and the innermost (interface) Mg layer. 
The ratio of the probability density on the surface relative to the 
interface layer of the high- (respectively, low-) energy state SS tend to 
increase (decrease) with increasing film thickness, and is $\sim1.3$ 
($\sim0.85$) in the 12-layer case. 
In the presence of the W substrate, some of the quantum-well states, and in 
particular those in the range [$E_F - 6$ eV, $E_F - 3$ eV] 
\cite{Schiller04,Kwak96}, remain fully localized within the Mg films---similar 
to the confined quantum-well states obtained from {\it ab initio} calculations 
in Ag films on Fe \cite{Wei03}, or sharp Fano resonances,  located mostly 
within the Mg film. 
Other Mg quantum-well states, instead, and in particular the states 
corresponding to the feature which crosses $E_F$, and is 
indicated by the arrow in Fig.~\ref{fig:ldos}, become broader resonances, 
corresponding to quantum-well states displaying an increased probability 
density within the W substrate.  The resonant character of the quantum-well 
states which cross the Fermi energy tends to smoothen the jump in 
$\lambda$ observed for thicknesses between 7  and 9 layers.

Hence, in spite of the presence of the surface/interface states, and the 
resonant character of the quantum-well states  crossing the Fermi 
energy, the qualitative behavior of $\lambda$, inferred from the 
particle-in-a-box model, is fully supported by the {\it ab initio} 
calculations. 
The variation in the decay length we predict from the {\it ab initio} 
calculations is expected to have a direct, exponential impact of the 
electron transfer rate by tunneling---from the metal to the $O_2$ 
molecule---which has been proposed to control the initial sticking of the 
oxygen molecules impinging on the surface, via the attractive image 
charge potential on the ionized O$_2^{-}$ molecule \cite{Hellman03}. 
Assuming a transfer rate by tunneling proportional to 
$e^{-d/\lambda}$, with $d$ the distance between the metal surface and the 
center of mass of the molecule, and considering, e.g.,  a distance 
$d \approx 3.5$ \AA---within the expected physisorption range of the 
O$_2$ molecules \cite{Hellman03,Bungaro97}, a 10 \% variation in 
$\lambda$ produces a 100 \% change in the transfer rate. Such a 
charge is of the order of magnitude of the experimental change in the 
oxidation rate at low O$_2$ exposure \cite{Aballe04}. 

The peak positions of the calculated DOS in the energy range 
[$E_F - 4$ eV, $E_F$], in Fig.~\ref{fig:ldos}, compare well with the  
near-normal-incidence PE measurements \cite{Aballe04}, except for a 
systematic shift, by about +2 Mg ML, of the theoretical spectra 
with respect to the measured spectra. In our calculations, the first 
quantum-well state with energy higher than the surface/interface states 
SS, enters the occupied-state spectrum at a Mg thickness of 8-9 ML. 
In the experiment, this state is found, instead,  to enter the 
valence-band spectrum at a nominal thickness of 6-7 ML. 
The same shift is observed between the calculated maximum/minima of  
$\lambda$ and the experimental maximum/minima of the reactivity. Such a 
shift is probably due to the use of a commensurate,  
dislocation-free interface atomic structure, in the calculations, 
to describe a heavily lattice-mismatched epitaxial system.  
The presence of strained Mg layers or dislocations at the interface 
could result in the entrance of a quantum-well state at a smaller 
value of the number of layers of the Mg film.  

The changes in $\lambda$ we predict should be observable, e.g., by STM near 
terrace steps on a Mg surface characterized by regions of different film 
thicknesses \cite{Aballe04}. Also, He scattering experiments have 
indicated an apparent step-height oscillation in the layer-by-layer growth 
of Pb on Ge due to quantum-size effects \cite{Crottini97}. Such changes 
were recently shown to be due mostly to a displacement of the topmost layer 
charge density \cite{Morgante03}, in contrast to previous interpretations  
in terms of a displacement of the surface atomic planes. 
This recent analysis corroborates thus the present finding on the behavior 
of the decay length of the Mg films. We note that the changes in the decay 
length of the density of states near the Fermi energy may also be related 
to a recent observation of quantum-size effects on the chemisorption 
properties of Cu(001) thin films \cite{Danese04}. 
In addition, layer-KKR (Korringa-Kohn-Rostoker) calculations \cite{Rous99}  
have indicated that the lifetime of negative ionic states of molecules, 
adsorbed on supported metal thin films varies with the thickness of 
the film, through coupling to quantum-well states. This effect 
(not yet measured, to our knowledge) was associated, in the calculations, with 
oscillations in the amplitude, at the position of the molecule, of the 
density of empty states above $E_F$. This calculated behavior, consistent 
with the trend we find for $\lambda$, also supports the interpretation we 
propose for the reactivity changes. 

In conclusion, on the basis of  {\it ab initio} calculations for 
Mg(0001) ultrathin films on W(110), we have shown that the decay length 
in vacuum of the electronic local density of states at the Fermi energy 
exhibits substantial oscillations with the film thickness. The decay length 
is maximal, as a function of film thickness, when a quantum-well state 
passes through the Fermi energy. 
The changes we predict in the decay length are expected to have 
a major impact on the tunneling rate in the electron transfer 
mechanism, which is believed to control the initial sticking of O$_2$ on the 
Mg surface. We therefore propose that the experimental  tuning 
of the reactivity is due to quantum oscillations in the electronic 
density decay length, which should be observable by STM. 

We thank L. Aballe, A. Barinov, M. Kiskinova, N. Stojic, and G. Trimarchi 
for helpful discussions. The computations have been performed
using the PWscf package \cite{pwscf}. We acknowledge support by the INFM 
within the framework of the ``Iniziativa Trasversale Calcolo Parallelo''.


\begin{thebibliography}{99}
\bibitem{nous} For  5 to 12 Mg ML on W, the relaxation of 
the topmost interlayer spacing (the largest relaxation) was found 
to be in the range +1.3 to +1.5\ \%.
\bibitem{nousw} We employed symmetric slabs with less vacuum, using the 
electrostatic potential as a reference in the vacuum.
\end{thebibliography}
\end{document}